\documentclass[letterpaper,english,reprint,aps,prl,superscriptaddress,showpacs,longbibliography]{revtex4-1}
\usepackage[latin9]{inputenc}
\usepackage{color}
\usepackage{textcomp}
\usepackage{amsmath}
\usepackage{amssymb}
\usepackage{graphicx}
\usepackage{gensymb}
\usepackage{setspace}

\usepackage[colorlinks,linkcolor=red,anchorcolor=black,citecolor=blue]
 {hyperref}

\makeatletter

\pdfpageheight\paperheight
\pdfpagewidth\paperwidth

\DeclareFontEncoding{LGR}{}{}

\ProvideTextCommand{\~}{LGR}[1]{\char126#1}

\makeatother

\begin{document}
\title{Highly efficient storage of 25-dimensional photonic qudit in a cold-atom-based quantum memory}
\author{Ming-Xin Dong}

\affiliation{Key Laboratory of Quantum Information, University of Science and Technology
of China, Hefei, Anhui 230026, China.}
\affiliation{Synergetic Innovation Center of Quantum Information and Quantum Physics,
University of Science and Technology of China, Hefei, Anhui 230026,
China.}
\affiliation{School of Physics and Materials Engineering, Hefei Normal University,
Hefei, Anhui 230601, China.}
\author{Wei-Hang Zhang}
\affiliation{Key Laboratory of Quantum Information, University of Science and Technology
of China, Hefei, Anhui 230026, China.}
\affiliation{Synergetic Innovation Center of Quantum Information and Quantum Physics,
University of Science and Technology of China, Hefei, Anhui 230026,
China.}
\author{Lei Zeng}
\affiliation{Key Laboratory of Quantum Information, University of Science and Technology
of China, Hefei, Anhui 230026, China.}
\affiliation{Synergetic Innovation Center of Quantum Information and Quantum Physics,
University of Science and Technology of China, Hefei, Anhui 230026,
China.}
\author{Ying-Hao Ye}
\affiliation{Key Laboratory of Quantum Information, University of Science and Technology
of China, Hefei, Anhui 230026, China.}
\affiliation{Synergetic Innovation Center of Quantum Information and Quantum Physics,
University of Science and Technology of China, Hefei, Anhui 230026,
China.}

\author{Da-Chuang Li}
\email{dachuangli@ustc.edu.cn}
\affiliation{School of Physics and Materials Engineering, Hefei Normal University,
Hefei, Anhui 230601, China.}

\author{Guang-Can Guo}
\affiliation{Key Laboratory of Quantum Information, University of Science and Technology
of China, Hefei, Anhui 230026, China.}
\affiliation{Synergetic Innovation Center of Quantum Information and Quantum Physics,
University of Science and Technology of China, Hefei, Anhui 230026,
China.}

\author{Dong-Sheng Ding}
\email{dds@ustc.edu.cn}

\affiliation{Key Laboratory of Quantum Information, University of Science and Technology
of China, Hefei, Anhui 230026, China.}
\affiliation{Synergetic Innovation Center of Quantum Information and Quantum Physics,
University of Science and Technology of China, Hefei, Anhui 230026,
China.}

\author{Bao-Sen Shi}
\email{drshi@ustc.edu.cn}

\affiliation{Key Laboratory of Quantum Information, University of Science and Technology
of China, Hefei, Anhui 230026, China.}
\affiliation{Synergetic Innovation Center of Quantum Information and Quantum Physics,
University of Science and Technology of China, Hefei, Anhui 230026,
China.}
\date{\today}

\begin{abstract}
Building an efficient quantum memory in high-dimensional Hilbert spaces is one of the fundamental requirements for establishing high-dimensional quantum repeaters, where it offers many advantages over  two-dimensional quantum systems, such as a larger information capacity and enhanced noise resilience.~To date, there have been no reports about how to achieve an efficient high-dimensional quantum memory.~Here,~we experimentally realize a quantum memory that is operational in Hilbert spaces of up to 25 dimensions with a storage efficiency of close to 60\%.~The proposed approach exploits the spatial-mode-independent interaction between atoms and photons which are encoded in transverse size-invariant orbital angular momentum modes. In particular, our memory features uniform storage efficiency and low cross-talk disturbance for 25 individual spatial modes of photons, thus allowing storing arbitrary qudit states programmed from 25 eigenstates within the high-dimensional Hilbert spaces, and eventually contributing to the storage of a 25-dimensional qudit state.~These results would have great prospects for the implementation of long-distance high-dimensional quantum networks and quantum information processing.
\end{abstract}

\maketitle

\textsl{Introduction}.\rule[2.2pt]{0.3cm}{0.04em}Quantum memories \cite{lvovsky2009optical,sangouard2011quantum} that enable quantum state storage and its on-demand retrieval are essential
requirements for quantum-repeater-based quantum communication networks \cite{briegel1998quantum,kimble2008quantum} and scalable quantum computation \cite{kok2007linear}. The storage efficiency exceeding the 50\% threshold \cite{vernaz2018highly,wang2019efficient,guo2019high} is necessary for practical applications due to the fundamental requirements of beating the quantum no-cloning limit without post-selection \cite{grosshans2001quantum} or realizing error correction in linear optical quantum computation
\cite{varnava2006loss}. Although  quantum memory has been widely demonstrated
in conventional two-dimensional (or qubit) quantum systems, it is highly
desirable to realize a high-dimensional quantum memory since manipulating a
photon in a high-dimensional Hilbert space, i.e., qudit, provides
many advantages over the qubit systems in terms of practical quantum
information processing. For example, qudits enable networks to carry more information and increase their channel
capacity via superdense coding in quantum communication \cite{erhard2018twisted,erhard2020advances,krenn2014generation}; for quantum cryptography, it has been shown that qudits can provide
a more secure flux of information against eavesdroppers \cite{bechmann2000quantum,bechmann2000quantumprl,cerf2002security,walborn2006quantum}
since the upper bound of limited cloning fidelity, given by $F_{{\rm clon}}^{d}=1/2+1/(d+1)$,
scales inversely with the dimension \cite{erhard2018twisted}, and they also
feature a better resilience to noise \cite{sheridan2010security,ecker2019overcoming}.
Moreover, qudit systems allow the simplification of quantum logic
gates \cite{lanyon2009simplifying}, and permit the enhanced
fault tolerance \cite{campbell2014enhanced} as well as the efficient distillation
of resource states \cite{campbell2012magic} in quantum computation.
In this regard, the capability to sufficiently store the qudit resources with high efficiency is of crucial importance for constituting high-dimensional networks so as to distribute high-capacity 
information in long-distance quantum communication and facilitate the complex
quantum computation.

Qubit memories have been widely demonstrated in many schemes that
usually encode photons in polarization \cite{ding2015raman,vernaz2018highly,wang2019efficient,gundougan2012quantum,PhysRevLett.111.240503} degree of freedom (DOF). However, such DOF can only support the two-dimensional encodings
involved with the quantum memory operation. To build up a qudit memory that can store high-dimensional information, alternative DOFs, such as which-path \cite{chou2005measurement,moehring2007entanglement,choi2008mapping,pu2017experimental,PhysRevLett.119.130505},
and time-bin
\cite{simon2007quantum,saglamyurek2011broadband,clausen2011quantum}, have been proposed in a variety of physical systems.~In addition, the photonic
transverse spatial mode, e.g., orbital angular momentum
(OAM) mode  \cite{krenn2017orbital,franke2017optical,ding2013single,nicolas2014quantum,parigi2015storage,ding2015quantum,zhang2016experimental,wang2021efficient,Ye2022long}, has attracted rapidly growing interest because of its advance of inherent infinite dimensionality. The storage of these spatial qutrit states with an efficiency of 20\%
using the electromagnetically induced transparency (EIT) scheme \cite{ding2014toward}
and efficiency of approximately 30\% through the off-resonant Raman protocol \cite{zhang2016experimental,reim2010towards,reim2011single} have been reported.
However, to date, the maximum available dimensionality of quantum memory in experiment is limited to \textit{d}=3 and their efficiencies are far below the
50\% threshold, largely limiting their practical applications in quantum
information processing.~The implementation of quantum memories
both having high efficiency and supporting high dimensions
is highly desirable but remains an open challenge.

\begin{figure}[t]
\includegraphics[width=1\columnwidth]{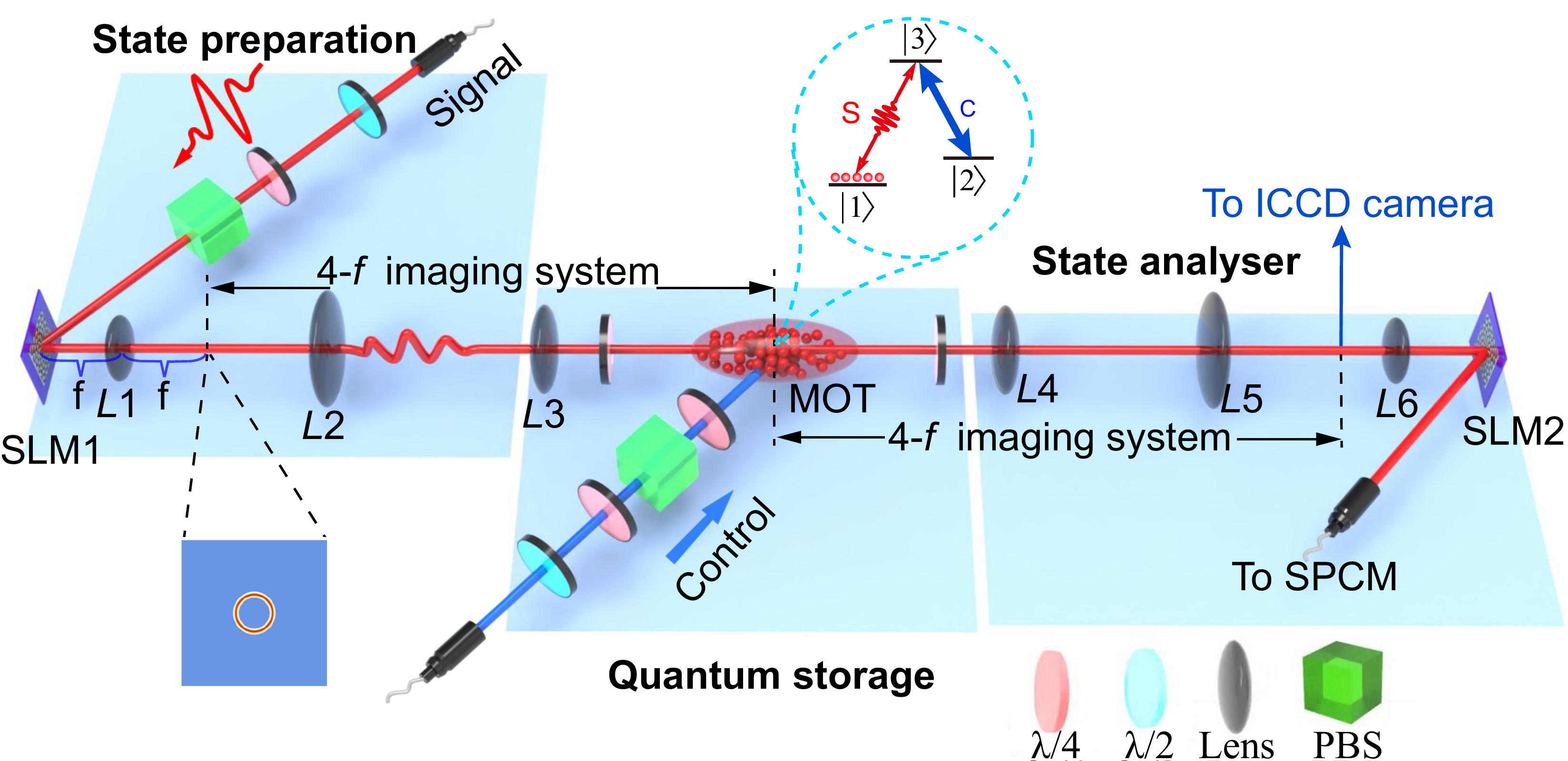} \caption{Schematic
experimental setup. The qudit signal, encoded in POV
mode via SLM 1 and lens \textit{L}1, is mapped into the atomic ensemble
for subsequent storage. Here, the signal and control fields are both circularly polarized ($\sigma^{+}$), and
the control field is beam expanded to have a waist of 4 mm to completely
cover the signal field at the centre of medium.}
\label{setup}
\end{figure}

There are two main challenges to realizing efficient high-dimensional
quantum memories.~The first is to establish a uniform
light-matter interface to achieve identical efficiencies for different
spatial modes.~The imbalanced storage efficiencies in storing different
spatial modes will significantly degrade the storage fidelity of the
qudit state with the increase of dimensionality. Taking the experiment
using Laguerre-Gaussian (LG) mode as a case in point, the rapidly scaling of the mode waist in $\sqrt{m}$
($m$ is the number of modes) \cite{ding2015quantum} will lead
to significant differences in light-matter interactions for different
modes, thus largely limiting its applicability in higher-dimensional
quantum storage.~The second challenge is to constitute a highly efficient
storage medium capable of storing multiple modes as many as possible
\cite{grodecka2012high}.~To achieve this, one needs to take into
account several physical parameters simultaneously in the storage process,
including the transverse spatial extent of the storage medium, the waist
size of the input modes, and the optical depth (OD) of the medium \cite{wang2019efficient,cao2020efficient}. Therefore, the
uniform and efficient storage of a large number of modes is technically challenging.

Here, we demonstrate a high-dimensional quantum memory
working up to a 25-dimensional Hilbert space with a storage
efficiency of close to 60\%, using the EIT protocol \cite{chaneliere2005storage,eisaman2005electromagnetically,zhang2011preparation,dai2012holographic,chen2013coherent,hsiao2018highly}
in a laser-cooled atomic ensemble. Through constituting a highly efficient
spatial-mode-independent light-matter interface where photons
are encoded in a unique perfect optical vortex (POV) mode \cite{supple} with invariant
transverse size, we are able to store a 25-dimensional qudit by mapping
it onto the 25 balanced spatial modes at the centre of the storage medium,
and coherently retrieve these components with identical efficiencies
via a control laser. The demonstrated high-dimensional
quantum memory with high efficiency herein is promising for high-capacity
quantum communication and high-dimensional quantum information processing.

\textsl{Model and experimental setup.}\rule[2.2pt]{0.3cm}{0.04em}Our memory scheme based on spatial-mode-independent
light-matter interaction is involved with a three-level $\Lambda$-type
atomic system, where the signal field (with a Rabi frequency $\Omega_{{\rm p}}$) drives the level
$\left|{\rm 1}\right\rangle $ to $\left|{\rm 3}\right\rangle $ and
the control field (with a Rabi frequency $\Omega_{{\rm c}}$) drives the level $\left|{\rm 2}\right\rangle $
to $\left|{\rm 3}\right\rangle $ (Fig.~\ref{setup}, dashed circle).
The dynamical evolution of the probe field under the slowly-varying
envelope approximation can be described by the Maxwell equation as
follows:

\begin{equation}
\left[\frac{1}{c}\frac{\partial}{\partial t}+\frac{\partial}{\partial z}\right]\Omega_{{\rm p}}=i\frac{D_{e{\rm ff}}\Gamma}{2L}\sigma_{31}
\end{equation}
\\ where $\Gamma$ denotes the decay rate of
$\left|{\rm 3}\right\rangle $, $L$ is the length of medium, and $\sigma_{31}$
represents the atomic coherence between levels $\left|{\rm 1}\right\rangle $
and $\left|{\rm 3}\right\rangle $. $D_{e{\rm ff}}\propto N_{{\rm tr}}g_{31}L$
represents the effective OD of an atomic ensemble, where we
define an effective atomic density $N_{{\rm tr}}$ while considering
a structured light
field interacts with the storage medium in the transverse
orientation. $g_{31}$ represents the photon-atom coupling coefficient
between $\left|{\rm 1}\right\rangle $~and~$\left|{\rm 3}\right\rangle $.~It can be observed from Eq.~(1) that $D_{e{\rm ff}}$ significantly affects the performance
of storage, and we derive the numerical relation between the storage
efficiency~and~OD by solving the Maxwell-Bloch equations \cite{supple}.

For a spatial multi-mode quantum memory, it is necessary
to take into account the effective light-matter interaction volume
for different spatial modes.~Here, we focus on the coupling of the
structure field with the storage medium in the cross section,
because the transverse extent of the storage medium is a crucial parameter
in determining the capacity of multi-mode memory \cite{grodecka2012high}.
We assume the atomic ensemble with a Gaussian distribution of the
density in the radial direction $N_{{\rm tr}}(r)=N_{0}\exp[-r^{2}/(2\sigma_{r}^{2})]$.
$N_{0}$ refers to the mean atomic density, and $\sigma_{r}$ represents
the half width of the atomic ensemble \cite{supple}.
~In this work, we propose a scheme to establish
a uniform light-matter interface for the memory of a variety of modes
via interacting the photons encoded in POV mode
with the storage medium.~Theoretically, such spatial modes feature identical
transverse sizes for different $m$, and thus they are subject to
the same $N_{{\rm tr}}(r)$ of atoms when they undergo the storage
process. The interaction strength between the desired POV modes and medium
is uniform, which manifests as the same $D_{e{\rm ff}}$ and ultimately
contributes to the same storage efficiency for different $m$. Based
on this mechanism, we constitute a spatial-mode-independent quantum
memory for the further implementation of storage of high-dimensional
quantum states.
\begin{figure*}[t]
\includegraphics[width=1.9\columnwidth]{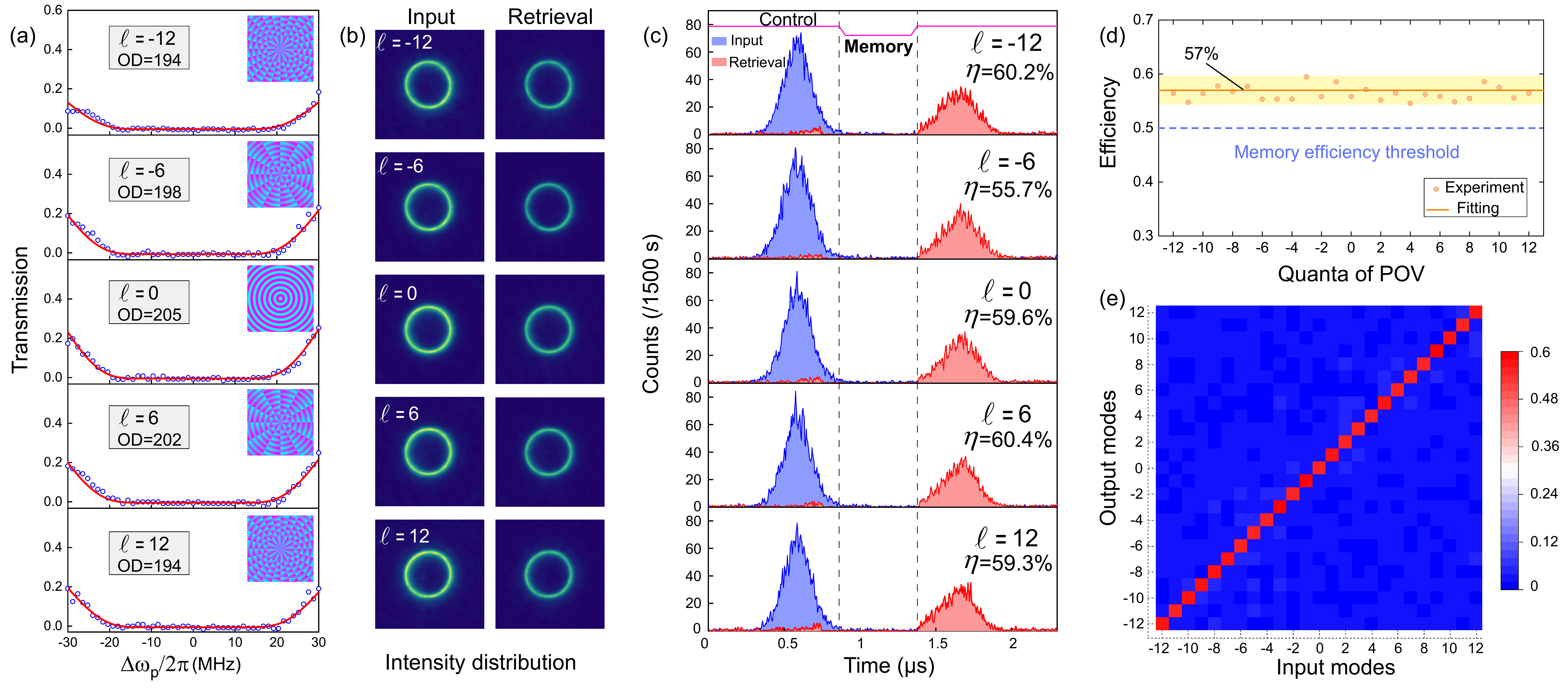} \caption{Performance of spatial multi-mode quantum storage. (a)
Measured absorption spectra for various spatial modes versus the signal detuning from the atomic resonance $\left|1\right\rangle \to\left|3\right\rangle $,
where the relative computer-controlled holograms loaded on the surface
of SLM1 are illustrated in the top right. (b) Transverse
intensity distributions of various modes recorded at the imaging
plane of the second 4-\textit{f} imaging system before (left) and
after (right) storage. (c) Temporal waveforms
of input (blue) and retrieved (red) pulses with temporal lengths of
about 500 ns for different modes. (d) Storage efficiencies versus the quanta of POV mode.
The shaded area represents the maximum fitted value that has been
expected, with a span of 1 sigma. (e) 25$\times$25 input-retrieved
cross-talk matrix formed by the basis set from $\ell=-12$ to 12.}
\label{distribution}
\end{figure*}

The experimental set-up for a high-dimensional quantum memory is schematically
depicted in Fig.~\ref{setup}. The qudits encoded in each spatial mode
are formed on the basis of the POV eigenstates $\left|\ell\right\rangle $
($\ell$ is chosen from -12 to 12), which is accomplished by means
of a Fourier transformation of the Bessel-Gaussian (B-G) state. In this
regard, we initially prepare the B-G states by projecting the attenuated
coherent states at the single-photon level onto a phase-only
spatial light modulator (SLM1) to shape the wave-fronts of photons
(Fig.~\ref{setup}, top left). The phase patterns loaded on
the SLM are programmed by a combination of Bessel and
Gaussian functions. Lens \textit{L}1
acting as a Fourier transformer is then used to transform the B-G
states to the POV states, which are subsequently mapped into the centre
of the atomic medium for storage with the assistance of a carefully aligned 4-\textit{f}
imaging system.

We next store and retrieve the POV states via the EIT storage protocol
in a rubidium medium. To ensure a high storage efficiency
of quantum memory, it is essential to prepare an optically thick atomic
ensemble with a large OD, which is implemented by using
a two-dimensional dark-line magneto-optical trap (MOT) technique in
our work. After a programmable storage
time, the signal photons are retrieved from the memory and sent into
a qudit state analyser, including the other 4-\textit{f} imaging system consisting
of lenses\textit{ L}4 and \textit{L}5, a Fourier lens \textit{L}6,
as well as a spatial-mode projector based on SLM2, a single-mode fiber
(SMF) and a single-photon counting module (SPCM), to fully characterize
the output states; see the right panel of Fig.~\ref{setup}.

\begin{figure}[t]
\includegraphics[width=1\columnwidth]{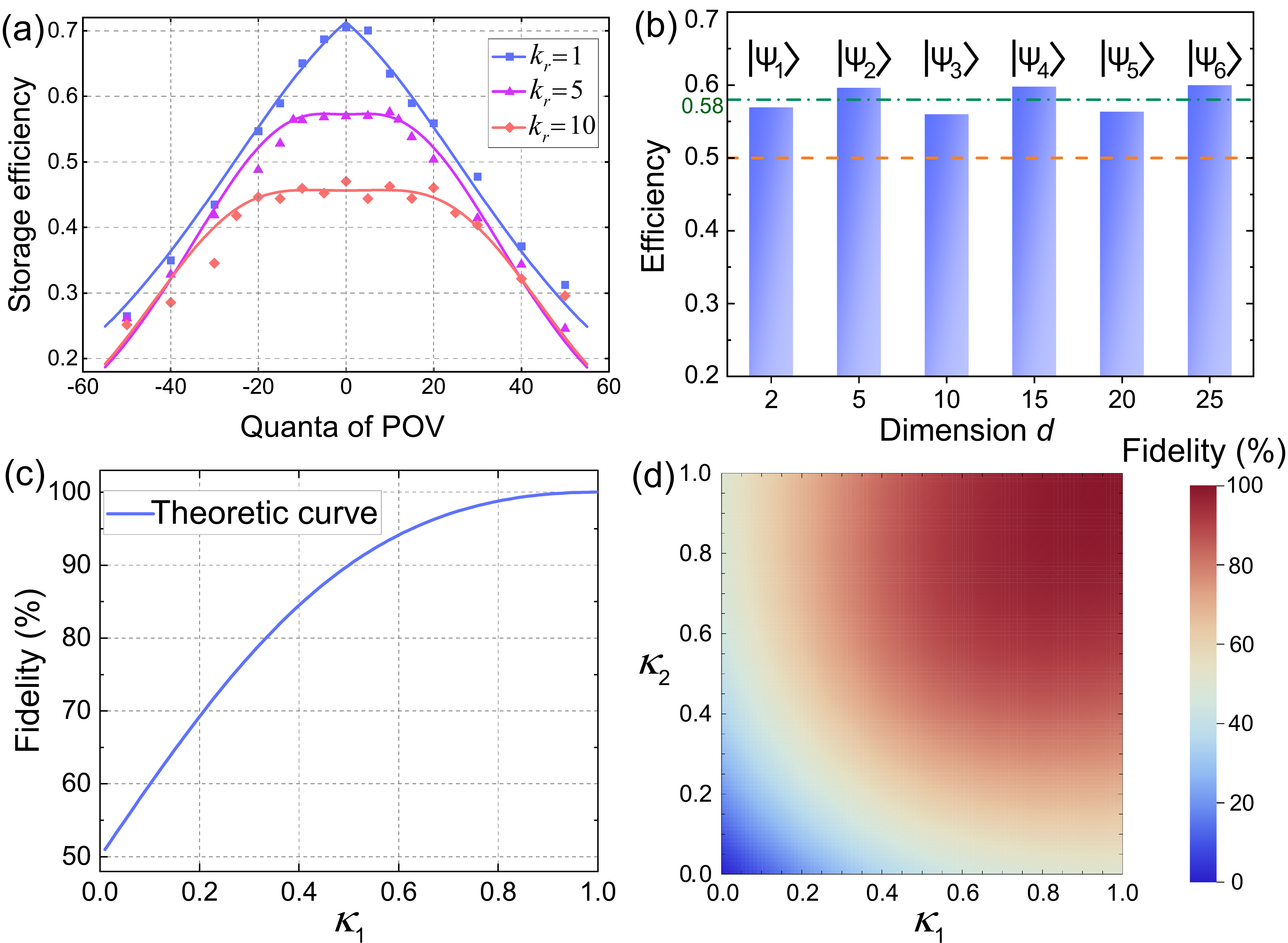}
\caption{Characteristics of high-dimensional storage. (a) Distributions of storage mode bandwidth for different radial wave vectors $k_{r}=1,5,10$, where the $k_{r}$ of 5 is used in the context. (b) Qudit
states with $d=2$, 5, 10, 15, 20 and 25 (see particular expressions in Ref \cite{supple}) versus storage efficiency.
(c) Numerical simulation of two-dimensional fidelity as a function of storage-efficiency-uniformity $\kappa_{1}$. (d) Theoretical analysis of fidelity versus $\kappa_{1}$ and $\kappa_{2}$ in the case of qudit with
$d=3$. }
\label{dimension}
\end{figure}

\textsl{Performance of multi-mode quantum memory.}\rule[2.2pt]{0.3cm}{0.04em}The key to achieving multi-mode storage in our scheme is to exploit the
mode-independent light-matter interaction. To confirm the accomplishment
of this particular photon-atom interface, we first measure the absorption
spectra for a variety of spatial modes, i.e. $\ell\in\left\{ -12,-6,0,6,12\right\} $
by scanning the detuning of signal from $-2{\rm \pi}\times30$ to
$+2{\rm \pi}\times30$ MHz, as depicted in Fig.~\ref{distribution}(a).~The nearly identical OD ($\sim$200) for various $\left|\ell\right\rangle $
indicates that the interactions between POV photons and atoms have
hardly any correlation with their mode number, thus allowing our memory
to be capable of carrying multiple spatial modes simultaneously. As shown in Fig.~\ref{distribution}(b), the spatial profiles of POV eigenstates with a mean
photon number\textit{ }of\textit{ n}=0.5 in the transverse orientation
are detected by an ICCD camera (iStar 334T series, Andor) working
at the single-photon level. The calculated high values \textsl{S} of similarity \cite{ding2013single} between input and retrieved states are 99.65\%, 99.63\%, 99.65\%, 99.61\% and 99.54\% for $\ell=-12,-6,0,6,12$ respectively, implying a faithful quantum storage for POV states.

Figure~\ref{distribution}(c) shows the temporal waveforms of the input
(blue) and retrieved pulses (red) after a one-pulse-delay storage time
for various spatial modes. As can be seen, the retrievals have almost
the same waveforms for different inputs, providing a clear evidence
that our memory exhibits identical characteristics for different
POV modes. To fully analyze the capacity of this spatial
multi-mode quantum memory, we investigate the memory efficiencies
of POV eigenstates across the entire range (from -12 to 12) with a step
of $\Delta\ell=1$; see Fig.~\ref{distribution}(d).~Their approximately
the same values at around 57\% clearly illustrate that our memory
enables 25 spatial-mode storage with efficiency beyond 50\%. Note
that the overall storage-efficiency distributions for different radial wave vector $k_{r}$ \cite{supple} in a wider mode range
are shown in Fig.~\ref{dimension}(a). Figure
\ref{distribution}(e) gives the experimental cross-talk between the 25 orthogonal
bases after retrieval. The average contrast \cite{supple}, given by \textsl{}$C=1/25\sum\nolimits _{m}C_{m}$, is estimated to be 92.4$\pm$1.6\%,
thereby revealing a low overlap noise between orthogonal spatial modes.

\begin{figure}[t]
\includegraphics[width=1\columnwidth]{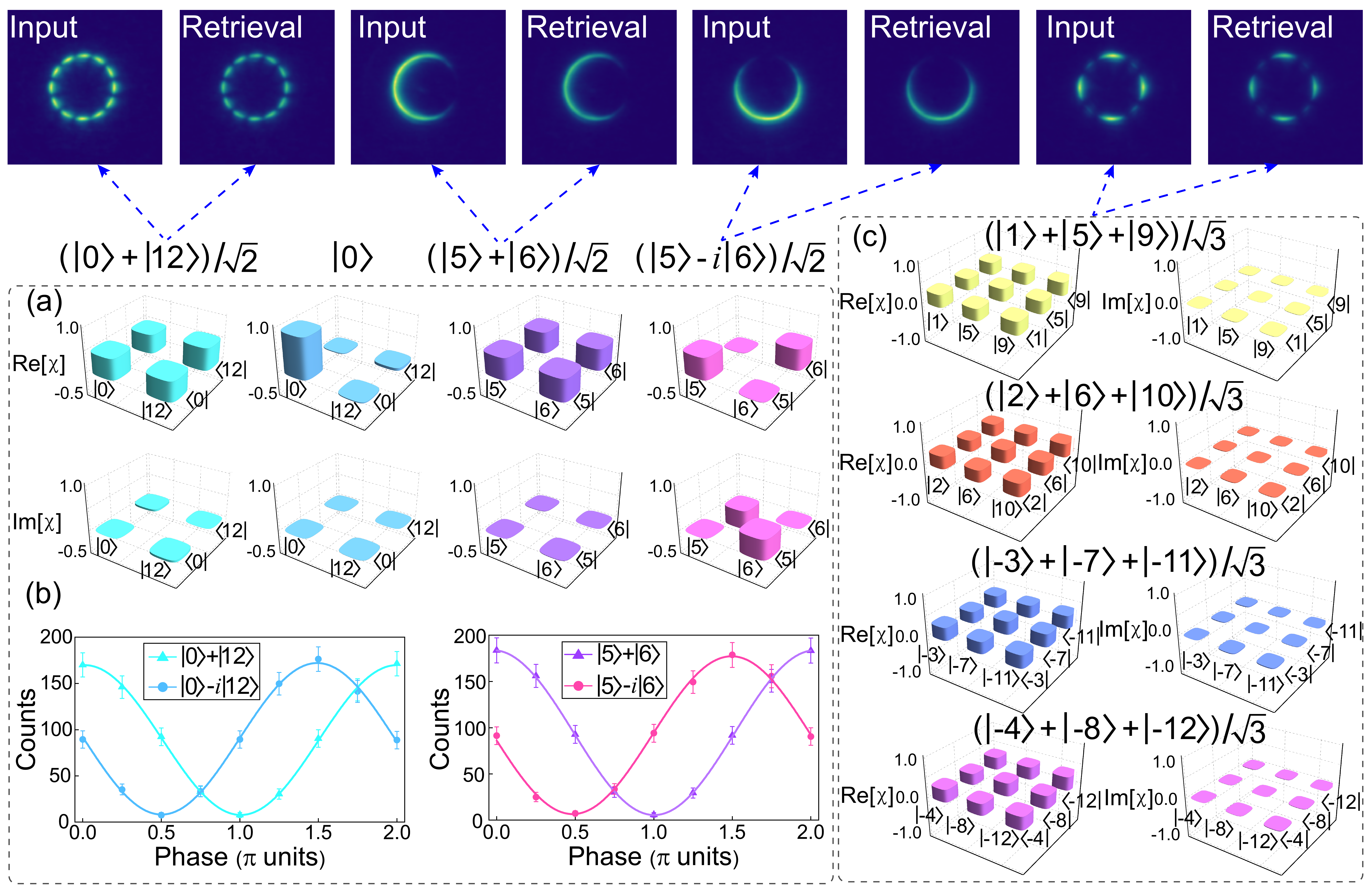} \caption{Demonstration of the storage of quantum states programmed
by arbitrary quanta. (a) Reconstructed real and imaginary
parts of density matrices of the retrieved arbitrary quantum states
with $d$=2 in different subspaces.~(b) Single-photon
interference fringes for different states. (c) Reconstructed density matrices of the retrieved qudits with $d$=3
in arbitrarily selected subspaces. The upper panel illustrates the
spatial profiles of the corresponding quantum states before and after
storage. The mean number of photons per pulse here is $n=0.5$.}
\label{programmable}
\end{figure}

In multi-mode memory, the uniform storage efficiency for each POV eigenstate plays a crucial role in high-dimensional
storage.~We consider a high-dimensional quantum superposition
state with the dimensionality of $d$, i.e.~the so-called qudit state
$\left|{\rm \psi}\right\rangle _{{\rm Input}}=1/\sqrt{d}\left(\left|\ell_{1}\right\rangle +\left|\ell_{2}\right\rangle +\cdots+\left|\ell_{d}\right\rangle \right)$
as input.~The retrieved state after storage can be written as

\begin{spacing}{0.5}
\[
\left|\psi\right\rangle _{{\rm Retrieval}}=1/\sqrt{\sum\nolimits _{m=1}^{d}\eta_{m}^{2}}(\eta_{1}\left|\ell_{1}\right\rangle +\eta_{2}\left|\ell_{2}\right\rangle +\cdots
\]

\end{spacing}

\begin{spacing}{0.75}
\noindent
\begin{equation}
+\eta_{d}\left|\ell_{d}\right\rangle )\ \ \ \ \ \ \ \ \ \ \ \ \ \ \ \ \ \ \ \ \
\end{equation}

\end{spacing}

\noindent where $\eta_{1},\cdots,\eta_{d}$ denote the storage efficiency
for the corresponding eigenmodes. $\left|\psi\right\rangle _{{\rm Retrieval}}$
can be further simplified to $\eta/\sqrt{d}\left(\left|\ell_{1}\right\rangle +\left|\ell_{2}\right\rangle +\cdots+\left|\ell_{d}\right\rangle \right)$
if $\eta_{1},\cdots,\eta_{d}$ are all equal to a constant represented
by $\eta$. In this case, the storage efficiency of the qudits has no
dependence on the dimensionality \textit{$d$}, as displayed by the
results in Fig.~\ref{dimension}(b).
Thus, our memory allows storing arbitrarily dimensional qudits with the same efficiency even
when $d$ is up to 25.

Storage fidelity is a critical performance parameter that has to be
taken into account in quantum memory. For the storage of qudit in
terms of multiple spatial modes, its fidelity is extremely sensitive
to the uniformity of the storage efficiency for the internal orthogonal
states. For simplicity, we consider the case of quantum states with
\textsl{$d$} = 2, as shown in Fig.~\ref{dimension}(c), where a parameter
$\kappa_{1}$ is defined as the ratio of storage efficiencies between
$\left|\ell_{1}\right\rangle $ and $\left|\ell_{2}\right\rangle $,
i.e.~$\kappa_{1}=\eta_{2}/\eta_{1}$. It can be found that the imbalanced
atomic storage ($\kappa_{1}\ll1)$ would largely reduce the fidelity, as estimated by the formula $F=\left[{\rm Tr}\left(\sqrt{\sqrt{\rho_{{\rm T}}}\rho_{{\rm retrieval}}\sqrt{\rho_{{\rm T}}}}\right)\right]^{2}$,
where $\rho_{{\rm T}}$ and $\rho_{{\rm retrieval}}$ represent the
density matrices corresponding to the target and retrieval states.
In Fig.~\ref{programmable}(a),
we reconstruct the retrieved density matrices using the quantum state tomography (QST) method for a set of qubit states
constituted by arbitrary
eigenstates (e.g. $\left|0\right\rangle $,
$\left|12\right\rangle $, $\left|5\right\rangle $, $\left|6\right\rangle $
are chosen herein) after storage. The average fidelity of 95.8\%
without any corrections
is in good agreement with the theoretical expectation, and the measured
single-photon interference fringes [Fig.~\ref{programmable}(b)] with
an average visibility of 92.3\% demonstrate that
the coherence between two components of the qubits is well preserved during
storage.

\begin{figure}[t]
\includegraphics[width=1.0\columnwidth]{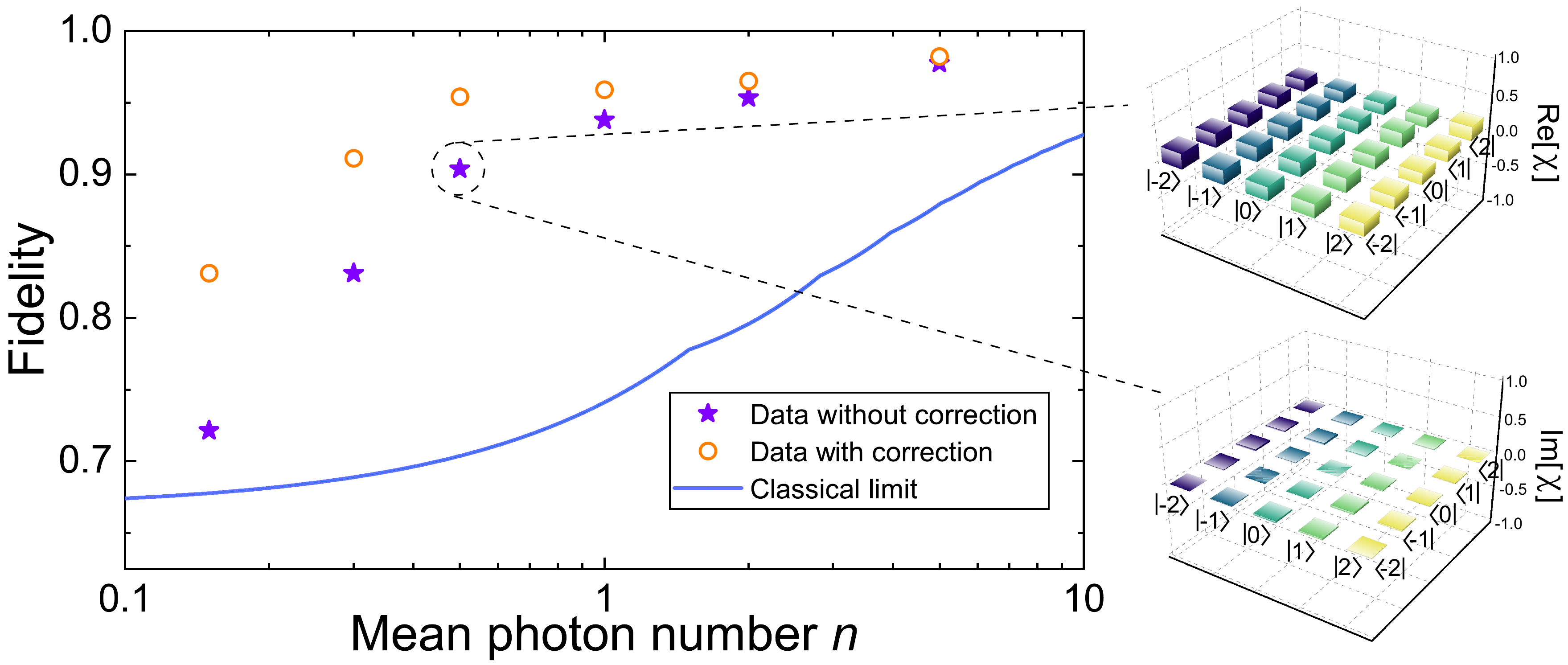} \caption{Storage in high-dimensional space exceeding the classical
benchmark.~The measured fidelities as a function of the mean photon
number per pulse \textsl{n}. The purple/yellow points are experimental
data without/with background subtraction. The blue solid line is the
classical limit after considering the finite storage efficiency and
Poissonian statistics of the input. }
\label{mean photon}
\end{figure}

In analogy to the case of \textsl{$d$} = 2, Fig.~\ref{dimension}(d)
illustrates the effect of efficiency-uniformity between internal modes on
the fidelity for \textsl{$d$} = 3, where $\kappa_{2}$ is defined
as $\eta_{3}/\eta_{1}$.
To obtain a high fidelity, $\kappa_{1}$
and $\kappa_{2}$ should both approach unity. In Fig.~\ref{programmable}(c),
we randomly choose three eigenvectors in the range from $\left|-12\right\rangle $
to $\left|12\right\rangle $ to prepare the high-dimensional states for storage.
The high mean fidelity is measured to be 96.4\% owing to $\kappa_{1}\approx\kappa_{2}\approx1$.
~Note that these results can hardly be obtained
in those experiments \cite{ding2015quantum} using conventional vortex
modes (e.g., LG mode) because of the inevitable non-uniform efficiency
for different spatial modes. Moreover, we characterize the retrieved
state of $\left|{\rm \psi}_{2}\right\rangle $ for \textsl{$d$} =5, and the raw fidelity reaches $90.7\pm0.7\%$
(the error bar is estimated from Poissonian statistics and
using Monte Carlo simulations),
as shown in the right panel of Fig.~\ref{mean photon}.~All these
experimental results indicate our memory capability of storing arbitrary-mode-encoded
qudit states programmed from 25 eigenvectors.

To further prove the quantum nature of the memory, we compare
the fidelities obtained in our experiment with the maximum available
fidelities in a classical memory device based on a completely classical
strategy \cite{gundougan2012quantum,nicolas2014quantum,parigi2015storage,vernaz2018highly}.~After considering the Poissonian statistics of photon number for a
coherent state, the classical fidelity threshold for a state with
a fixed photon number \textsl{N} can be written as
\begin{equation}
F_{{\rm class}}(n)=\sum\limits _{N=1}^{\infty}\left(\frac{N+1}{N+2}\right)\frac{e^{-n}n^{N}}{(1-e^{-n})N!}
\label{eq:eq2}
\end{equation}

\noindent where $n$ is the mean photon number per pulse.
As presented in Fig.~\ref{mean photon}, the solid line
is the theoretically classical limit after taking $\eta=0.57$ in
our work. We observe that all the experimental points exceed the classical
benchmark for different mean photon numbers, which confirms the quantum
character of our device.

\begin{figure}[t]
\includegraphics[width=1.0\columnwidth]{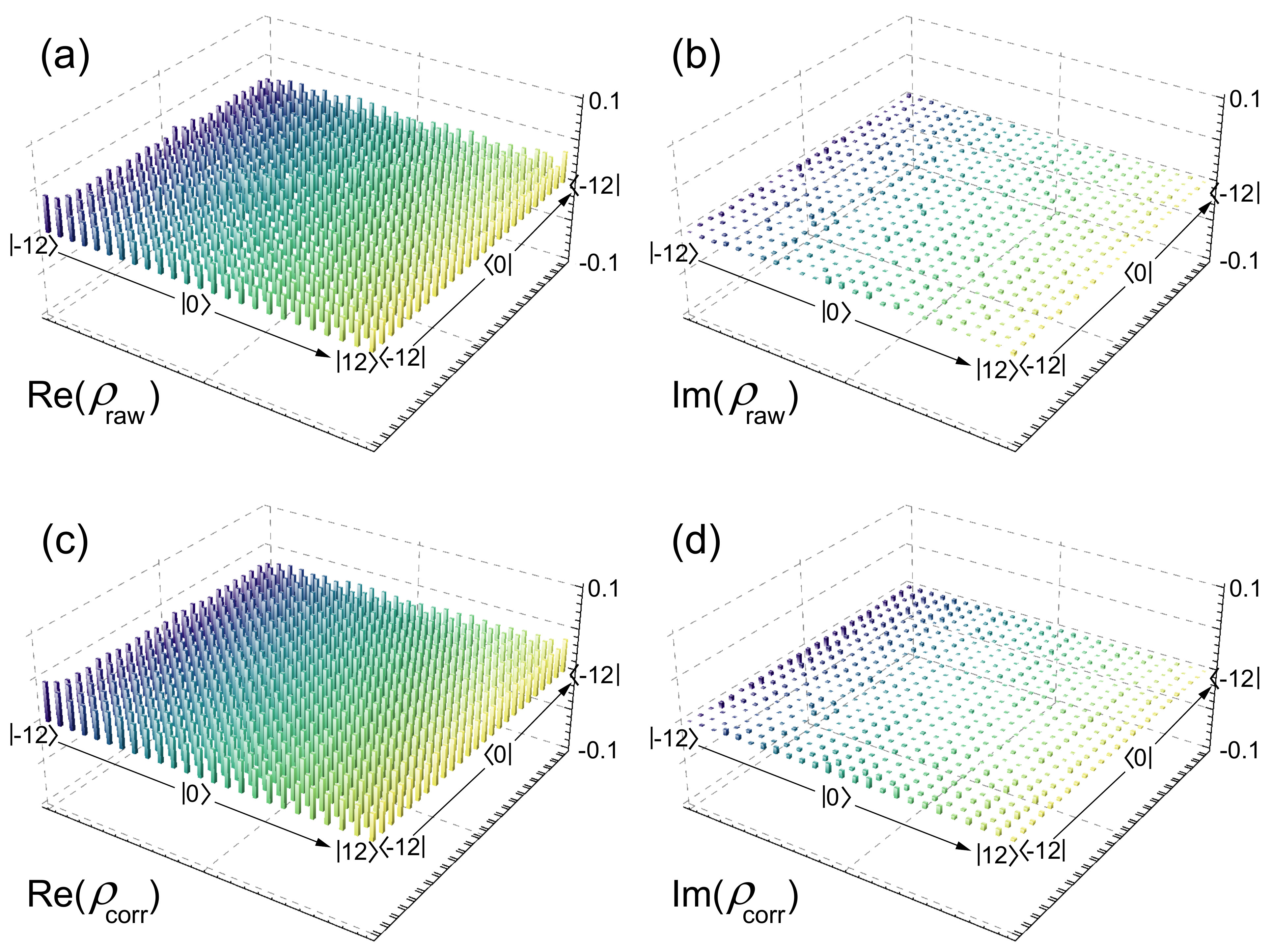} \caption{Experimental realization of 25-dimensional qudit storage.
The characterization of the retrieved qudit state $\left|{\rm \psi}_{6}\right\rangle $
after the storage process by performing QST. (a)/(c) and (b)/(d) are
the real and imaginary parts of the reconstructed density matrices for
retrieved state $\left|{\rm \psi}_{6}\right\rangle $ without/with
background correction, respectively.}
\label{25 dimension}
\end{figure}

We now turn to study the capability of our memory to store a 25-dimensional
quantum state.
The main challenge to achieving the storage of a 25-dimensional qudit state is to preserve
the identical memory efficiency for each mode, thus preventing the
decay of coherence between 25 spatial modes during the storage process.
Here, a 25-dimensional qudit state $\left|\Psi\right\rangle $
 given by a coherent superposition of 25 individual
spatial modes from $\left|{\rm -12}\right\rangle $ to $\left|{\rm 12}\right\rangle $
is prepared for the demonstration of 25-dimensional qudit storage, which is represented as
\begin{equation}
\left|\Psi\right\rangle =\frac{1}{\sqrt{25}}\sum\limits _{\ell=-12}^{+12}\left|\ell\right\rangle
\end{equation}
\indent To fully characterize the retrieved state, we perform the high-dimensional
QST \cite{supple,thew2002qudit}, where
the real and imaginary parts of the reconstructed density matrix without
(with) background correction are plotted in the logical basis of \{$\left|{\rm -12}\right\rangle ,$$\left|{\rm -11}\right\rangle $,
$\left|{\rm -10}\right\rangle $, $\cdots$, $\left|{\rm 12}\right\rangle $\},
as shown in Fig.~\ref{25 dimension}(a,b) (Fig.~\ref{25 dimension}(c,d)), respectively. The
raw fidelity between the retrieved states and ideal state is estimated
to be $72.8\pm0.6\%$, where the imperfection fidelity is
mainly caused by the dark counts of the detector and residual control
laser leakage. After the subtraction of the background, the fidelity reaches $90.3\pm0.6\%$, far exceeding the classical limit of 70.2\%
for mean photon number $n=0.5$, where the memory
efficiency of state $\left|{\rm \psi}_{6}\right\rangle $ equals to
60\% is taken into account. Note that the residual
fidelity is primarily due to the imperfections in the qudit preparation
and measurement. All the above results clearly beat the classical
benchmark, thus demonstrating the quantum character of our 25-dimensional
memory implementation.

\textsl{Conclusion.}\rule[2.2pt]{0.3cm}{0.04em}In summary, we have experimentally demonstrated the efficient quantum storage for high-dimensional quantum states with
\textsl{d} up to 25 using the POV modes of photons.~The reported high-dimensional
quantum memory achieves a storage efficiency of \textgreater 50\%, exceeding
the threshold value for practical quantum information applications.
Remarkably, the dimensionality of this memory is scalable
to as high as 100 through further optimization of the waist of POV modes
\cite{supple}, thus presenting a clear route to the scalability of
dimensions. In addition, our multi-mode memory is also promising
for the compatibility with fiber-based quantum information transfer
systems,
~which are capable of spatially-structured photon transmission \cite{liu2020multidimensional,cao2020distribution}.
The high-dimensional quantum memory demonstrated herein
gives a great perspective for the practical high-capacity and long-distance
quantum communication networks.

  This work was supported by National Key R\&D Program of China (Grants No. 2017YFA0304800), Anhui Initiative in Quantum Information Technologies (Grant No.~AHY020200), the National Natural Science Foundation of China (Grants No. U20A20218, No. 61722510, No. 11934013, No. 11604322, No. 12204461), and the Innovation Fund from CAS, and the Youth Innovation Promotion Association of CAS under Grant No. 2018490.

\bibliographystyle{Plain}

\end{document}


\title{Supplemental Material for Highly efficient storage of 25-dimensional photonic qudit in a cold-atom-based quantum memory}
\author{Ming-Xin Dong}
\affiliation{Key Laboratory of Quantum Information, University of Science and Technology
of China, Hefei, Anhui 230026, China.}
\affiliation{Synergetic Innovation Center of Quantum Information and Quantum Physics,
University of Science and Technology of China, Hefei, Anhui 230026,
China.}
\affiliation{School of Physics and Materials Engineering, Hefei Normal University,
Hefei, Anhui 230601, China.}
\author{Wei-Hang Zhang}
\affiliation{Key Laboratory of Quantum Information, University of Science and Technology
of China, Hefei, Anhui 230026, China.}
\affiliation{Synergetic Innovation Center of Quantum Information and Quantum Physics,
University of Science and Technology of China, Hefei, Anhui 230026,
China.}
\author{Lei Zeng}
\affiliation{Key Laboratory of Quantum Information, University of Science and Technology
of China, Hefei, Anhui 230026, China.}
\affiliation{Synergetic Innovation Center of Quantum Information and Quantum Physics,
University of Science and Technology of China, Hefei, Anhui 230026,
China.}
\author{Ying-Hao Ye}
\affiliation{Key Laboratory of Quantum Information, University of Science and Technology
of China, Hefei, Anhui 230026, China.}
\affiliation{Synergetic Innovation Center of Quantum Information and Quantum Physics,
University of Science and Technology of China, Hefei, Anhui 230026,
China.}
\author{Da-Chuang Li}
\email{dachuangli@ustc.edu.cn}
\affiliation{School of Physics and Materials Engineering, Hefei Normal University,
Hefei, Anhui 230601, China.}
\author{Guang-Can Guo}
\affiliation{Key Laboratory of Quantum Information, University of Science and Technology
of China, Hefei, Anhui 230026, China.}
\affiliation{Synergetic Innovation Center of Quantum Information and Quantum Physics,
University of Science and Technology of China, Hefei, Anhui 230026,
China.}
\author{Dong-Sheng Ding}
\email{dds@ustc.edu.cn}

\affiliation{Key Laboratory of Quantum Information, University of Science and Technology
of China, Hefei, Anhui 230026, China.}
\affiliation{Synergetic Innovation Center of Quantum Information and Quantum Physics,
University of Science and Technology of China, Hefei, Anhui 230026,
China.}
\author{Bao-Sen Shi}
\email{drshi@ustc.edu.cn}

\affiliation{Key Laboratory of Quantum Information, University of Science and Technology
of China, Hefei, Anhui 230026, China.}
\affiliation{Synergetic Innovation Center of Quantum Information and Quantum Physics,
University of Science and Technology of China, Hefei, Anhui 230026,
China.}
\date{\today}
\maketitle
\section*{\mbox{\uppercase\expandafter{\romannumeral1}}. Experiment details}
As depicted in Fig.~1, the relevant quantum states $\left|{\rm 1}\right\rangle $, $\left|{\rm 2}\right\rangle $ and $\left|{\rm 3}\right\rangle $ correspond to the $^{{\rm 85}}{\rm Rb}$ atomic hyperfine levels $\left|5S_{1/2},F=2\right\rangle $, $\left|5S_{1/2},F=3\right\rangle $ and $\left|5P_{1/2},F=3\right\rangle $ respectively. Here, the POV signal is resonant with the atomic transition of $\left|{\rm 1}\right\rangle $ \ensuremath{\leftrightarrow} $\left|{\rm 3}\right\rangle $, and the auxiliary control field with Rabi frequency of $2\pi\times20.07$ MHz dynamically drives the transition of $\left|{\rm 2}\right\rangle $ \ensuremath{\leftrightarrow} $\left|{\rm 3}\right\rangle $ to achieve reversible transfer between photonic states and atomic collective spin excitation during the storage process.
The control field entering into the atomic medium has an angle of 3\degree{} with respect to the signal to write
and read the photonic qudit state. After an on-demand storage time, the retrieved state is characterized by a state analyser. The qudit analyser is set for implementing related
projective measurements of the retrieved states into distinct bases. Before
signal photons being measured by the detector, two homemade Fabry-Perot
etalons (with a total transmittance of 64\%, an isolation rate of 46 dB
and a bandwidth of 500 MHz), acting as frequency filters, are inserted
into the path (not shown in Fig.~1) to reduce the noise scattered
from the control field. The outputs are finally detected by the SPCM (avalanche
diode, PerkinElmer SPCM-AQR-16-FC; 60\% efficiency, maximum dark count
rate of 25/s) for recording the relative measurement counts.
\\\indent 4-\textit{f} imaging system.\rule[2.2pt]{0.3cm}{0.04em}The first 4-\textit{f} imaging system shown in Fig.~1 is
used to map the photonic POV modes onto the centre of the storage medium,
and mainly consists of lenses \textit{L}2 and \textit{L}3,
which are separated by a distance of $f_{2}+f_{3}$. Here, $f_{2}$
and $f_{3}$ refer to the focal lengths of lenses \textit{L}2 and \textit{L}3. The focal lengths of lenses \textit{L}1, \textit{L}2, \textit{L}3, \textit{L}4, \textit{L}5 and \textit{L}6 are 75, 500, 300, 300, 500 and 75 mm, respectively.
The POV spatial structure is located in the front focal plane of \textit{L}2,
and its image is obtained at the back focal plane of \textit{L}3\textit{.
}In addition, this shrunken 4-\textit{f} imaging system is further exploited
to reduce the waist of the input photonic modes, with a scaling
ratio of $f_{3}$/$f_{2}$, to better match with the transverse size of
the storage medium.
\\\indent High-$d$ state expressions.\rule[2.2pt]{0.3cm}{0.04em}In Fig.~3(b) of main text, the used six quantum states are $\left|{\rm \psi}_{1}\right\rangle =\left(\left|0\right\rangle +\left|12\right\rangle \right)/\sqrt{2}$, $\left|{\rm \psi}_{2}\right\rangle =1/\sqrt{5}\sum\nolimits _{\ell=-2}^{\ell=+2}\left|\ell\right\rangle $, $\left|{\rm \psi}_{3}\right\rangle =1/\sqrt{10}\left(\sum\nolimits _{\ell=-5}^{\ell=-1}\left|\ell\right\rangle +\sum\nolimits _{\ell=+1}^{\ell=+5}\left|\ell\right\rangle \right)$, $\left|{\rm \psi}_{4}\right\rangle =1/\sqrt{15}\sum\nolimits _{\ell=-7}^{\ell=+7}\left|\ell\right\rangle $, $\left|{\rm \psi}_{5}\right\rangle =1/\sqrt{20}(\sum\nolimits _{\ell=-10}^{\ell=-1}\left|\ell\right\rangle +$ $\sum\nolimits _{\ell=+1}^{\ell=+10}\left|\ell\right\rangle $), $\left|{\rm \psi}_{6}\right\rangle =1/\sqrt{25}\sum\nolimits _{\ell=-12}^{\ell=+12}\left|\ell\right\rangle $, respectively.

\section*{\mbox{\uppercase\expandafter{\romannumeral2}}. High-optical-depth cold atomic ensemble}

We trap the cold atoms of Rubidium 85 (\textsuperscript{{\small{}85}}Rb)
in a two-dimensional dark-line MOT. Each trapping
laser beam has a power of 36 mW with a beam waist of 2 cm. Two vertically oriented repump laser beams have a total power of 60 mW with a beam waist of 2 cm, with two copper bars located in their central positions to prepare
two dark-line images \cite{zhang2012dark,wang2019efficient}. These
images are overlapped at the centre of the MOT along the longitudinal
axis by using two 4-\textit{f} imaging
systems, and thus constituting a dark-line volume. Here, the diameters of the copper bars are 1.5 mm. The
experimental repetition rate is 100 Hz, and the experimental operation
window is 1.3 ms, during which the MOT magnetic field is switched
off completely. Initially, all of the atoms are prepared in the ground state $\left|{\rm 1}\right\rangle $
by turning off the repump laser 500 \textmu s earlier than the trapping laser. Thanks to the high
OD herein, we have achieved a storage efficiency of 72.3\%
for a single photon in Gaussian mode, as shown in Fig. S1.
\\\indent Figure S1 shows the temporal waveforms of input (blue) and retrieved
pulses (red) after a one-pulse-delay storage time for Gaussian mode.
The storage efficiency of quantum memory can be defined as

\begin{equation}
\eta=\frac{\int\left|\psi_{{\rm retrieval}}(t)\right|^{2}dt}{\int\left|\psi_{{\rm input}}(t)\right|^{2}dt}
\end{equation}
where $\psi_{{\rm input}}(t)$ and $\psi_{{\rm retrival}}(t$) represent
the input and retrieved quantum states. The storage efficiency for the Gaussian mode is estimated to be 72.3\% by integrating
the counts of photons in their entire temporal waveforms.

\renewcommand{\thefigure}{S1}
\begin{figure}[t]
\includegraphics[width=0.65\columnwidth]{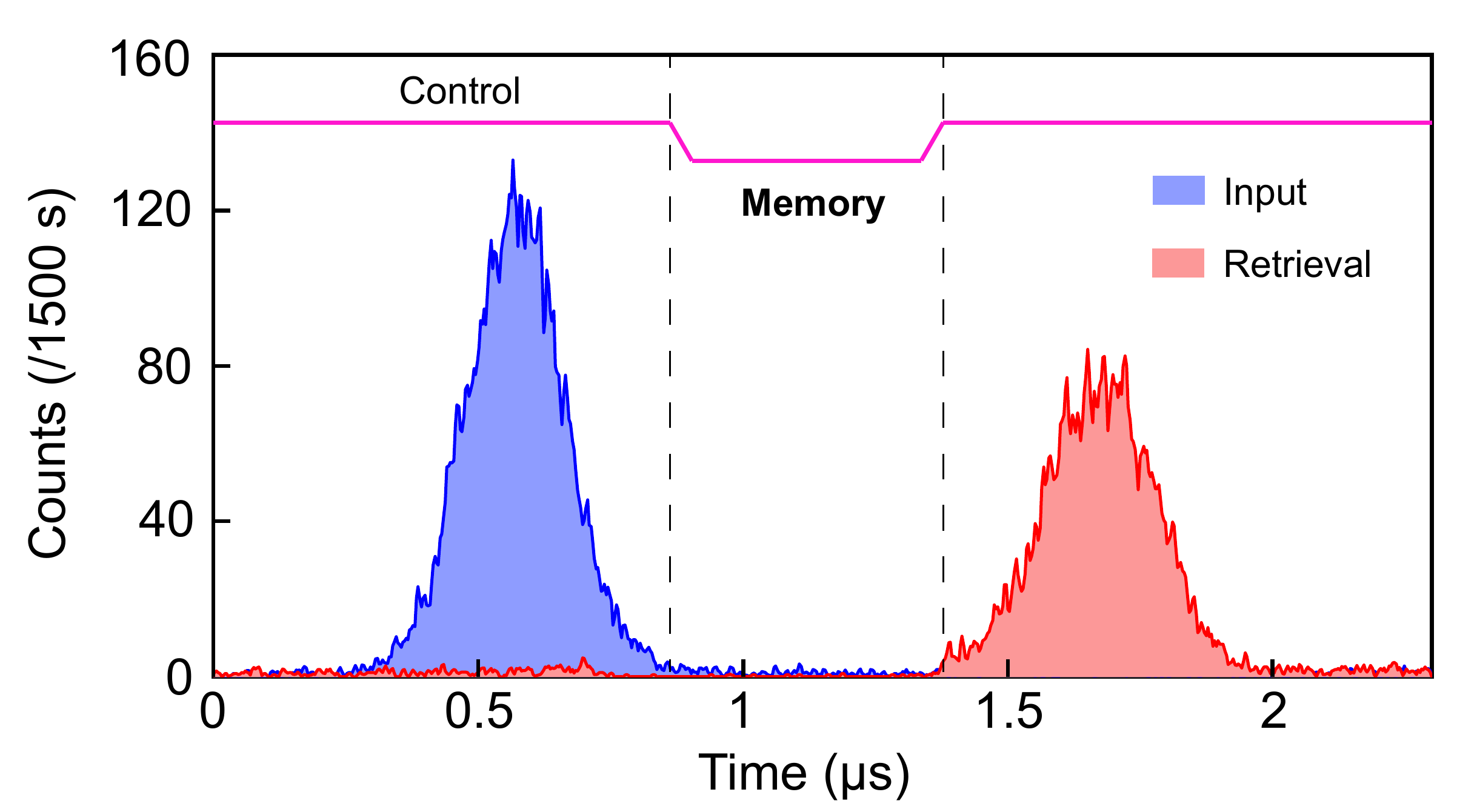} \caption{The temporal waveform of input (blue) and retrieved (red) pulses for
Gaussian mode. }
\label{setup}
\end{figure}

\section*{\mbox{\uppercase\expandafter{\romannumeral3}}. Preparation and analysis of POV modes}

The structured light beams carrying OAM are
of many interests due to their wide applications in micromanipulation
\cite{grier2003revolution}, optical imaging \cite{uribe2013object}
and quantum information processing \cite{ding2013single,nicolas2014quantum}.
The photons carrying OAM are described by a helical phase factor
$e^{i\ell\phi}$, with $\ell$ being the topological charge indicating
the OAM of $\ell\hbar$ per photon and $\phi$ is the azimuthal angle. Here,
$\ell$ takes any integer values, therefore the available Hilbert-space
dimension is infinite in principle, which provides a huge information
capacity for classical and quantum information systems. However, the
conventional optical vortices, e.g., LG beams,
always exhibit a strong dependence of their transverse size on the
topological charge number, thus limiting their further applications
in some circumstances, such as optical trapping and tweezing, multiplexed
optical communication using a single fibre with a fixed annular index
profile \cite{yan2012free}, as well as the high-dimensional quantum information
processing \cite{zhang2016experimental}. To overcome these limitations,
the concept of a perfect optical vortex \cite{vaity2015perfect,liu2021broadband}
has been proposed whose radius is independent of $\ell$. The POV mode is defined as the Fourier transformation of a B-G function, and the complex field amplitude is given by

\begin{equation}
E_{{\rm POV}}^{\ell}(r,\phi)=i^{\ell-1}\frac{\omega_{g}}{\omega_{0}}\exp(i\ell\phi)\times\text{exp}(-\frac{r^{2}+r_{r}^{2}}{\omega_{0}^{2}})I_{\ell}(\frac{2r_{r}r}{\omega_{0}^{2}})
\end{equation}
\\ where $\omega_{g}$ is the beam waist of the Gaussian mode, $\omega_{0}{\rm =2}f/k\omega_{g}$
denotes the beam waist of the Gaussian mode at the focal plane with a wave vector
$k{\rm =}2\pi/\lambda$, in which $f$ is the focal length of the Fourier
lens. $r_{r}{\rm =}k_{r}f/k$ is the radius of POV mode ($k_{r}$
is the radial wave vector). For the case of large $k_{r}$ and small
$\omega_{0}$, Eq.~(4) can be further reduced to
\begin{equation}
E_{{\rm POV}}\left(r,\phi\right)\propto i^{\ell-1}/k_{r}\delta\left(r-r_{r}\right)\exp\left(i\ell\phi\right)
\end{equation}
where $\delta(r)$ represents the Dirac
delta function.
We can clearly observe that this ideal POV mode has a transverse radius
of $r_{r}$ which is independent of $\ell$. In the experiments, one
can obtain the POV mode at the Fourier plane of the B-G phase. In addition, the radius of the perfect vortex could be controlled
by varying the radial wave vector $k_{r}$ of the B-G beam.

\section*{\mbox{\uppercase\expandafter{\romannumeral4}}. THEORETICAL ANALYSIS OF THE MODE-INDEPENDENT QUANTUM STORAGE}

We consider a three-level atom model in the storage process, and
the dynamics of the laser-driven atomic system can be described by
the master equation as follows
\begin{equation}
\frac{\partial\hat{\rho}}{\partial t}=-\frac{i}{\hbar}[H_{{\rm int}},\hat{\rho}]-\frac{1}{2}\left\{ \hat{\Gamma},\hat{\rho}\right\}
\end{equation}
 where $H_{{\rm int}}$ is the interaction Hamiltonian that describes
the light-atom coupling, and the second term attributes to the atomic
relaxation that describes the radiative decay and the decoherence
processes of the excited state and ground state. Under the rotating-wave
approximation \cite{fleischhauer2005electromagnetically}, $H_{{\rm int}}$
is given by

\begin{equation}
H_{\mathop{{\rm int}}}=-\frac{\hbar}{2}\left[\begin{array}{ccc}
0 & 0 & \Omega_{{\rm p}}\\
0 & -2(\Delta\omega_{{\rm p}}-\Delta\omega_{{\rm c}}) & \Omega_{c}\\
\Omega_{{\rm p}} & \Omega_{c} & -2\Delta\omega_{{\rm p}}
\end{array}\right]
\end{equation}

\noindent \\where $\Omega_{{\rm p}(c)}$ and $\Delta\omega_{{\rm p(c)}}$
denote the Rabi frequency and detuning of the signal (control) field,
respectively. The Maxwell-Bloch equations can then be written as

\begin{equation}
\begin{array}{l}
\partial_{t}\sigma_{31}=(i\Delta\omega_{{\rm p}}-\gamma_{31})\sigma_{31}+\frac{i}{2}\Omega_{{\rm c}}\sigma_{21}+\frac{i}{2}\Omega_{{\rm p}}
\\[3mm]
\partial_{t}\sigma_{21}=\left[i(\Delta\omega_{{\rm p}}-\Delta\omega_{{\rm c}})-\gamma_{21}\right]\sigma_{21}+\frac{i}{2}\Omega_{c}\sigma_{31}
\\[3mm]
(1/c\partial_{t}+\partial_{z})\Omega_{{\rm p}}=i\frac{D_{{\rm eff}}\Gamma}{2L}\sigma_{31}
\end{array}
\end{equation}

 Here, $\sigma_{ij}$ represents the atomic coherence between
levels $\left|i\right\rangle $ and $\left|j\right\rangle $. As referred
to Ref.~\cite{hsiao2018highly}, we can obtain the numerical relation
between the effective atomic optical depth $D_{{\rm eff}}$ and storage efficiency $\eta$
by taking the Fourier transform, which is given by

\begin{equation}
\eta=\frac{\exp(-2\gamma_{21}D_{{\rm eff}}\Gamma/\Omega_{c}^{2})}{\sqrt{1+32\ln2\frac{\gamma_{31}D_{{\rm eff}}\Gamma}{\left(T_{{\rm p}}\Omega_{c}^{2}\right)^{2}}}}\times\frac{1}{2}\left[{\rm erf(2}\sqrt{\ln2}\kappa{\rm )+erf(2}\sqrt{\ln2}\frac{D_{{\rm eff}}\Gamma/(T_{{\rm p}}\Omega_{c}^{2})-\kappa}{\sqrt{1+32\ln2\frac{\gamma_{31}D_{{\rm eff}}\Gamma}{\left(T_{{\rm p}}\Omega_{c}^{2}\right)^{2}}}}{\rm )}\right]
\end{equation}

\noindent where $\gamma_{21}$ is the ground-state decoherence rate
between levels $\left|2\right\rangle $ and $\left|1\right\rangle $.
$\gamma_{31}=\Gamma/2$ is the decay rate of $\left|3\right\rangle \leftrightarrow$$\left|1\right\rangle $.
$T_{{\rm p}}$ denotes the full width at half maximum (FWHM) of signal
pulse duration. $\kappa$ is the proportionality of the time span, when the control field is switched off, to the $T_{{\rm p}}$.

\renewcommand{\thefigure}{S2}
\begin{figure}[t]
\includegraphics[width=0.8\columnwidth]{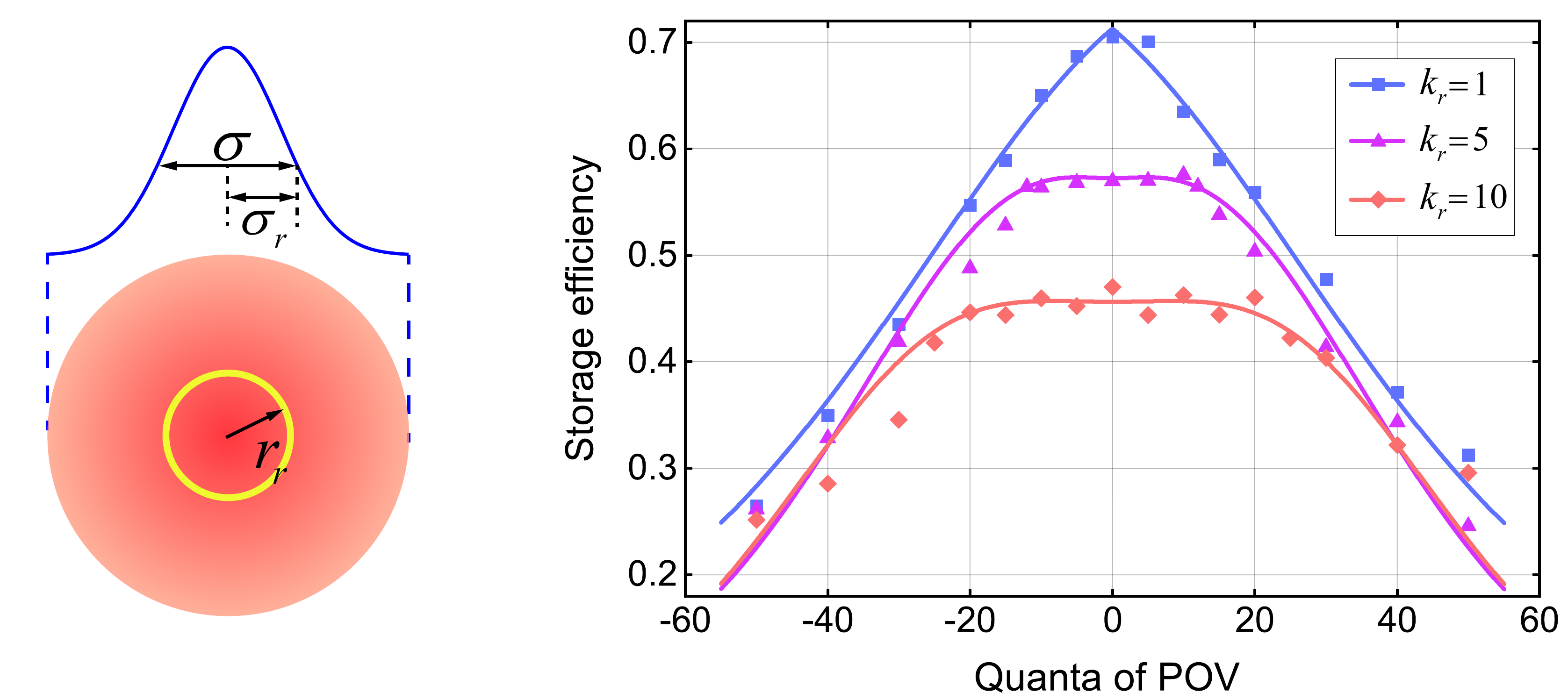} \caption{(left) Schematic of the cross-section distribution of the POV mode and
atomic ensemble with cylindrical symmetry. (Right) The measured storage
efficiency $\eta$ as a function of quanta of POV for several radial
wave vectors $k_{r}=1,5,10$. The symbols are experimental data and
the solid lines are the theoretically simulated curves. The fitting
parameters \{$T_{{\rm p}}$, $\sigma_{r}$, $\kappa$\} are \{300
ns, $0.275\pm0.025$ mm, 1.1\}, respectively. It can be seen that
the theoretical calculations are in good agreement with the experimental
results. }
\label{setup}
\end{figure}
\renewcommand{\thefigure}{S3}
\begin{figure}[t]
\includegraphics[width=0.9\columnwidth]{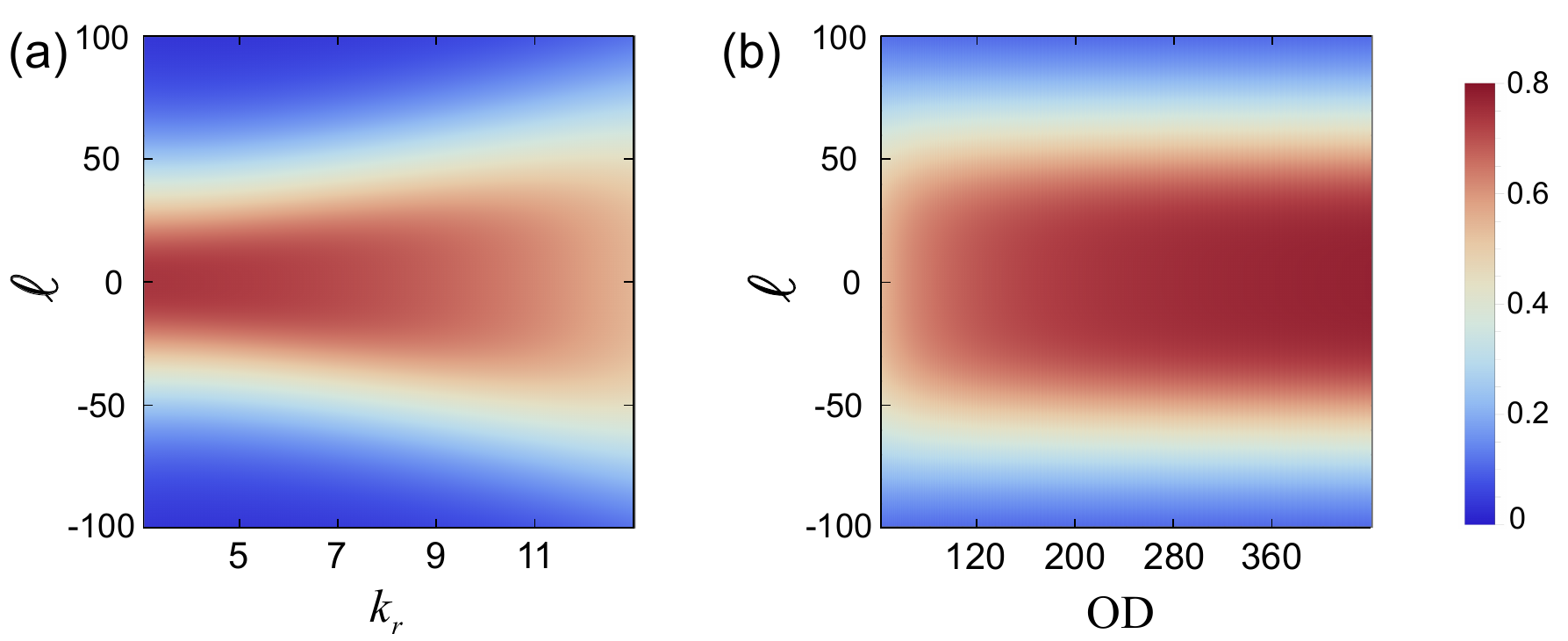} \caption{Theoretical simulations of the multi-mode storage performance. a, The distribution of storage efficiency $\eta$ as a function of $k_{r}$ and $\ell$. b, The storage efficiency $\eta$ versus OD and $\ell$ with $k_{r}=13$.}
\label{distribution}
\end{figure}
As depicted in the main text, we take into account that the atomic
ensemble has a Gaussian distribution of the density in the radial
direction $N_{{\rm tr}}(r)=N_{0}\exp[-r^{2}/(2\sigma_{r}^{2})]$ (Fig.~S2,
left). $r_{r}{\rm =}k_{r}f/k$ represents the transverse size of the POV
mode of signal. The same value of $r_{r}$ for different quanta $\ell$
results in a mode-independent light-matter interaction, which is
the key to realizing high-dimensional quantum storage in our work.

\section*{\mbox{\uppercase\expandafter{\romannumeral5}}. STORAGE-EFFICIENCY DISTRIBUTION FOR A WIDE MODE SPECTRUM}

The left panel of Fig.~S2 depicts the transverse distribution of POV
photons and atomic ensemble. The radius size of the POV mode with $k_{r}=5$
at the centre of the MOT is theoretically estimated to be $r_{r}=218$
\textmu m, which is very close to the measured value of 222 \textmu m
detected by the ICCD camera.

The POV mode can be derived from the Fourier transformation of a Bessel function.
As the generation of an ideal Bessel beam is difficult in the experiment,
we turn to its finite-energy approximation, i.e., the Bessel-Gaussian
beam, to prepare POV. When the quanta of POV is invloved in a large
range, it is worthwhile to consider a second-moment width of the beam
profile in the practical physical process, which is defined as \cite{pinnell2019perfect}:

\begin{equation}
\omega_{\ell}=\omega_{0}\sqrt{\ell+1}+r_{r}\sqrt{1+I_{\ell+1}(r_{r}^{2}/\omega_{0}^{2})/I_{\ell}(r_{r}^{2}/\omega_{0}^{2})}
\end{equation}

From this, we can find that the ratio of $r_{r}$ and $\omega_{0}$ is
a crucial parameter that could evaluate the quality of POV, where
$r_{r}/\omega_{0}\gg1$ is the ideal case. As shown in Fig.~S2 (right),
the storage efficiency of POV decreases as $\ell$ increases in the
case of large quanta, which agrees well with the theoretical result obtained
by combining Eqs.~(7) and (8). To achieve mode-independent
quantum storage over a wider range, one can control the Bessel parameters,
e.g. $k_{r}$, whereas the limitation of the radius size it causes in
the atomic ensemble should be taken into account. This difficulty can be overcome
by adjusting the scaling ratio of the controllable 4-\textit{f} imaging system used in our work. The capacity of
quantum memory allows for scaling in our scheme, which has a great
prospect to achieve a higher-dimensional quantum memory by means of
further optimization of several physical parameters, such as $k_{r}$
and OD. To clearly display it, we theoretically investigate these parameters for improving the storage efficiency and uniform-efficiency-mode range, as shown in Fig.~S3. The uniform-efficiency-mode range increase with the increase of $k_{r}$ whereas the uniform efficiency is decreased (Fig.~S3(a)). This limitation can be overcome by further improving the OD of the storage medium, as shown in Fig. S3(b). Therefore, we can achieve a quantum memory enabling higher mode and having higher storage efficiency based on our scheme.

\section*{\mbox{\uppercase\expandafter{\romannumeral6}}. Definitions of similarity and cross-talk}
To quantitatively assess the preservation of spatial structures for input and retrieved states, we calculate the similarity $S=\sum\nolimits _{i}\sum\nolimits _{j}A_{ij}B_{ij}/\sqrt{\left(\sum\nolimits _{i}\sum\nolimits _{j}A_{ij}^{2}\right)\left(\sum\nolimits _{i}\sum\nolimits _{j}B_{ij}^{2}\right)}$,
where \textit{A} and \textit{B }denote the grey-scale matrices of two images \cite{ding2013single}, and the subscripts $i$ and $j$ represent different pixels. The cross-talk disturbance in our work is quantified by defining a contrast $C_{m}=\left(E_{mm}-E_{mn,{\rm max}}\right)/E_{mm}$, where $m$, $n$ refer to the input and projected mode numbers chosen from -12 to 12, and $E_{mm}$, $E_{mn,{\rm max}}$ represent the diagonal coefficients and maximal off-diagonal coefficients inside the 25$\times$25 matrix [Fig.~2(e)], and the average contrast is estimated by \textsl{}$C=1/25\sum\nolimits _{m}C_{m}$.

\section*{\mbox{\uppercase\expandafter{\romannumeral7}}. Storage of coherent superposition states with specific modes}

We access the storage of coherent superposition states, $\left|\psi_{\pm\ell}\right\rangle =\left(\left|-\left|\ell\right|\right\rangle +\left|+\left|\ell\right|\right\rangle \right)/\sqrt{2}$, which are encoded in POV modes with quanta from low-order alphabets to high-order alphabets. The correspondingly coherent interference patterns are shown in Fig.~S4 when we chose $\ell$ from 1 to 12. The similarities between inputs (top) and outputs (bottom) are estimated to be 99.74\%, 99.64\%, 99.55\%, 99.70\%, 99.40\%, 99.70\%, 99.69\%, 99.70\%, 99.66\%, 99.60\%, 99.62\%, 99.63\%, indicating the high-fidelity storage for various superposition states. Meanwhile, the calculated storage efficiencies for these states are 60.6\%, 62.1\%, 62.1\%, 64.8\%, 62.2\%, 62.6\%, 60.9\%, 62.8\%, 61.2\%, 65.5\%, 60.7\%, 60.2\%, manifesting approximately the same value for different $\ell$, which agrees well with the theoretical expectation.
\renewcommand{\thefigure}{S4}
\begin{figure}[t]
\includegraphics[width=1\columnwidth]{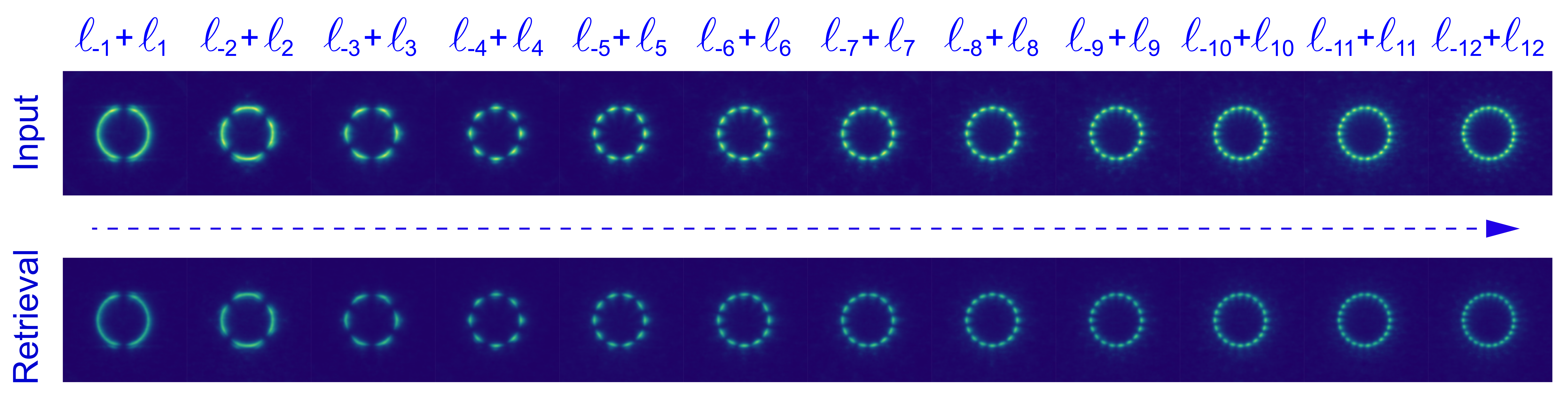} \caption{The intensity distributions for input (upper layer) and retrieved (lower layer) superposition states.}
\label{setup}
\end{figure}

\section*{\mbox{\uppercase\expandafter{\romannumeral8}}. Dimension versus storage time}

We have measured the storage time of memory for qudits with different dimensions, as shown in Fig.~S5. Here, the input qudit states of $\sum\nolimits _{\ell=-1}^{\ell=+1}\left|\ell\right\rangle /\sqrt{3}$, $\sum\nolimits _{\ell=-2}^{\ell=+2}\left|\ell\right\rangle /\sqrt{5}$, $\sum\nolimits _{\ell=-5}^{\ell=+5}\left|\ell\right\rangle /\sqrt{11}$, $\sum\nolimits _{\ell=-12}^{\ell=+12}\left|\ell\right\rangle /\sqrt{25}$ with dimensions of $d=3, 5, 11, 25$ respectively, are used to display the relation of storage time with dimensions.
It can easily be observed that our quantum memory is robust to the dimensions of quantum states since the storage time is almost the same for different dimensional qudits. Our memory features identical storage characters (e.g. storage efficiency and storage time) for 25 POV modes with l from -12, -11, -10 $\cdots$ 10, 11, 12. Thus, these same quantum storage characters for 25 individual eigenstates allow us to store arbitrary qudits with dimensions $d$ between 1 to 25, which shows the strong programmability of our memory.
In conclusion, it has great prospects for building a higher-dimensional quantum memory using our scheme.
\renewcommand{\thefigure}{S5}
\begin{figure}[b]
\includegraphics[width=0.6\columnwidth]{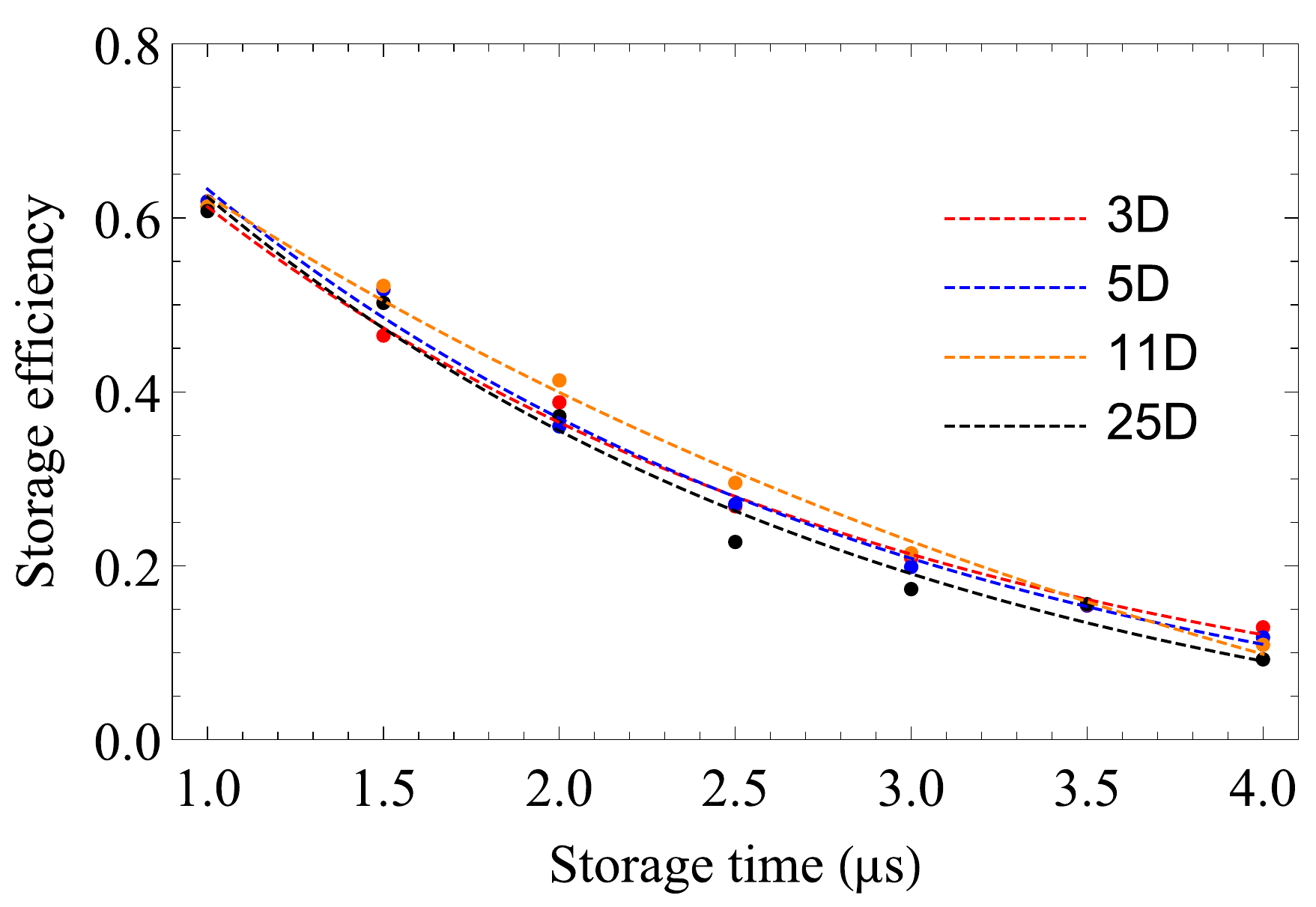} \caption{The storage time for qudits with different dimensions.}
\end{figure}

\section*{\mbox{\uppercase\expandafter{\romannumeral9}}. High-dimensional QST}
QST is an efficient and robust technique for accurately characterizing
the measured quantum states. Thus, to determine the full states of the retrieval in the storage process, we use a high-dimensional QST to reconstruct the density matrix. The \textit{d}-dimensional
density matrix $\rho_{d}$ can be described by
\begin{equation}
\hat{\rho}_{d}=\frac{1}{d}\sum\limits _{j=0}^{d^{2}-1}r_{j}\hat{\lambda}_{j}
\end{equation}

\noindent where $\hat{\lambda}_{j}$ represents the operator for a SU(\textit{d}) system, and $r_{j}=\left\langle \hat{\lambda}_{j}\right\rangle ={\rm Tr[}\hat{\rho}\hat{\lambda}_{j}{\rm ]}$
is the expectation value of the operators. In practical measurements,
an arbitrary but complete set of basis states $\left\{ \left|\psi_{i}\right\rangle \right\} $
is exploited to implement the related projection operations with operators
$\left\{ \hat{\mu}_{i}=\left|\psi_{i}\right\rangle \left\langle \psi_{i}\right|\right\} $.
Owing to the completeness of $\hat{\mu}_{i}$, this can be written as
$\hat{\mu}_{i}=\sum\nolimits _{j}A_{i}^{j}\hat{\lambda}_{j}$. Here,
we take the corresponding measurement values $n_{i}=N\left\langle \psi_{i}|\hat{\rho}|\psi_{i}\right\rangle =N{\rm Tr}[\hat{\rho}\hat{\mu}_{i}]{\rm =}N\sum\nolimits _{j}A_{i}^{j}{\rm Tr}[\hat{\rho}\hat{\lambda}_{j}]=N\sum\nolimits _{j}A_{i}^{j}r_{j}$
($N$ is a constant of proportionality, which is dependent on the detector efficiency
and count of photons). The density matrix can finally be written as
$\hat{\rho}_{d}=N^{-1}\sum\nolimits _{i,j}\left(A_{i}^{j}\right)^{-1}n_{i}\hat{\lambda}_{j}/d$.
Ultimately, we use the maximum-likelihood estimation technique with
the combination of 625 individually projected measurement values
to derive a physical density matrix for a 25-$d$ qudit, as shown in Fig. 6 of main text.

\bibliographystyle{Plain}